\documentclass{aa}
\usepackage{graphics}
\usepackage{times}

\begin{document}

\thesaurus{02(11.03.4 MG2016+112; 11.03.4 AXJ2019+112; 12.04.1; 12.07.1)}

\title{Weak lensing observations of the ``dark'' cluster MG2016$+$112}

\author{D. Clowe \inst{1} \and N. Trentham \inst{2} \and
J. Tonry\thanks{Visiting Astronomer at the W. M. Keck Observatory,
jointly operated by the California Institute of Technology and the
University of California} \inst{3}}

\offprints{D. Clowe}

\institute{Max-Planck-Institut fuer Astrophysik, Karl Schwarzschild
Str. 1, 85740 Garching, Germany \and Institute of Astronomy,
University of Cambridge, Madingley Road, Cambridge. CB3 0HA, United
Kingdom \and Institute for Astronomy, University of Hawaii, 2680
Woodlawn Drive, Honolulu, HI 96822, USA} 

\date{Received 10 January 2000 / Accepted 10 January 2000}

\maketitle

\begin{abstract}
We investigate the possible existence of a high-redshift ($z=1$)
cluster of galaxies associated with the QSO lens system MG2016$+$112.
From an ultra-deep $R$- and less deep $V$- and $I$-band Keck images
and a $K$-band mosaic from UKIRT, we detect ten
galaxies with colors consistent with the lensing galaxy within
225$h^{-1}$ kpc of the $z=1.01$ lensing galaxy.  This represents an
overdensity of more than ten times the number density of galaxies with
similar colors in the rest of the image.  We also find a group of
seven much fainter objects closely packed in a group only 27$h^{-1}$ kpc
north-west of the lensing galaxy.  We perform a weak
lensing analysis on faint galaxies in the $R$-band image and detect a
mass peak of a size similar to the mass inferred from X-ray
observations of the field, but located 64$\arcsec$
northwest of the lensing galaxy.  From the weak lensing data we rule
out a similar sized mass peak centered on the lensing galaxy at the
$2\sigma$ level.
\end{abstract}

\section{Introduction}

The QSO lens MG2016$+$112 is a long-standing enigma.  Spanning a few
arcseconds in size, the system contains two QSO images (designated as
components A and B, see Fig.~\ref{fig4}) and an extended structure
(designated C), all at redshift
$z=3.273$ (Lawrence et al.~\cite{Lawrence84}, Schneider et al.~
\cite{Schneider85}, \cite{Schneider86}).  The
extended structure appears as an arc in the infrared (Langston et al.~
\cite{Langston}; Lawrence et al.~\cite{Lawrence93};
Benitez et al.~\cite{Benitez}, hereafter B99), but consists of four
subcomponents in the radio, elongated 
along the arc (Garrett et al.~\cite{Garrett}).  Roughly in the middle
of the lens 
structure is a giant elliptical galaxy at redshift $z=1.01$
(designated D, Schneider et al.~\cite{Schneider85}), and another
galaxy of unknown  
redshift lies just outside the giant arc.  Various lensing models
have been applied to this system involving a single elliptical galaxy
as lens, a giant elliptical galaxy plus a cluster core lens, a giant
elliptical plus companion galaxy lens, and a double lens model
involving two galaxies at different redshift each lensing parts of the
system (Narasimha \&\ Chitre \cite{Narasimha}; Langston et
al.~\cite{Langston}; Nair \&\ Garrett \cite{Nair}; B99), but no model
has successfully reproduced all of the observed structures, positions,
and brightness ratios of the system.

An observation with the ASCA satellite revealed a strong X-ray source,
designated AXJ2019$+$112,
towards MG2016$+$112 with an emission line at 3.5 keV which could correspond
to FeXXV at $z=0.92$ or FeXXVI at $z=1.00$, the redshift of the
lensing galaxy (Hattori et al.~\cite{Hattori}, hereafter H97).  The
X-ray spectra results in 
an 8.6 keV gas temperature for a $z=1.0$ source, implying a massive
cluster ($M \approx 2\times 10^{14} h^{-1} M_\odot$ at $r = 250 h^{-1}$
kpc) at the redshift of the lens.
Infrared and optical observations of the field (Lawrence et
al.~\cite{Lawrence93}, Langston et al.~\cite{Langston},
Schneider et al.~\cite{Schneider85}) revealed no 
excess of galaxies around the giant elliptical lens, the supposed
brightest cluster galaxy (BCG), as would normally be found in a cluster.  
This led to the suggestion of a ``dark cluster'' associated with the
lens, in which a cluster-sized dark matter and hot gas overdensity
exists with few optically bright galaxies ($M/L_V > 2000
M_\odot/L_\odot$).  These observations were
even more puzzling as the observed iron line in the X-ray spectra
indicates a near-solar metallicity for the cluster, which implies a
long history of star formation in the region (H97).
A handful of galaxies at $z=1.01$ have been found spectroscopically
within a few arcminutes of the lens (Soucail et al.~\cite{Soucail}),
but these are not concentrated strongly around the lens.

Recently, B99 have found a slight overdensity of
red galaxies using deep $V$, $I$, and $K_s$ imaging in a region 200
kpc ($\sim $47\arcsec ) in radius around the QSO lens.  Assuming that
the galaxies they identified were the most luminous galaxies in a
Schechter cluster luminosity function, they calculated that the total
cluster luminosity was roughly ten times that of the BCG.  Thus the
proposed cluster has a mass-to-light ratio of only a few hundred
solar, typical of what is seen in other massive high-redshift clusters
(Luppino \&\ Kaiser \cite{Luppino}, Clowe et al.~\cite{Clowe98}).  The
observed galaxies are red in both $V-I$ and $I-K_s$, suggesting a old
stellar population 
and consistent with what is observed in other high-redshift clusters
(Trentham \&\ Mobasher \cite{Trentham98}, van Dokkum et al.~\cite{Dokkum}).  
These observations, however, are based on only nine
galaxies selected by two colors and confirmed by a third.  

In this paper we present deep optical and infrared images in $V$, $R$,
$I$, and $K$ of the MG2016$+$112 field with the goals of testing the
B99 selected galaxies with additional colors, detecting any
additional, fainter cluster galaxies, and measuring the weak lensing
signal in the field to confirm the presence and measure the mass of
any cluster.  In \S2 we discuss the images taken and the image
reduction process.  We discuss the colors and magnitudes we measure
for the proposed cluster galaxies of B99 as well as others
in the field in \S3.  Section 4 contains a weak lensing analysis of
the field.  We summarize and discuss our major conclusions in \S5.

Throughout this paper, unless stated otherwise, we assume $\Omega _0 =
1$, $\Lambda = 0$, and $H_0 = $100 km/s/Mpc.

\section{Observations and Data Reduction}
Deep $I$, $R$, and $V$-band imaging was performed on a field centered
on MG2016$+$112 using LRIS (Oke et al.~\cite{Oke}) in direct imaging mode at
the Keck II
telescope on the nights of 1998 July 22-23.  The resulting images
cover a 6\arcmin$\times$7\farcm 5 area in $I$ and $V$ and 7\farcm
5$\times$7\farcm 5 area in $R$ with 0\farcs 215 pixels.  The total
integration times are 1500s in $I$ and $V$ and 5400s in $R$, with
seeing of 0\farcs 50, 0\farcs 56, and 0\farcs 54 FWHM in $I$, $R$, and
$V$ respectively.

The individual images in each filter were de-biased using the overscan
strip and flattened with dome flats.  The $I$-band images were then
re-flattened with a medianed night-sky flat from a (relatively) empty
piece of sky taken during the same nights to remove fringing.  
The images were then remapped with a bi-cubic polynomial to correct
for focal plane curvature.  The polynomial for the mapping
was determined by requiring that the positions of the stars in each
remapped frame were consistent to within 0.1 pixels ($\approx 0\farcs
02$, the estimated rms error in the centroiding algorithm) rms with
each other and within the errors of the positions of the stars in a
Digital Sky Survey image and a catalog of stellar positions from the
United States Naval Observatory.  The mappings were
then checked using a second group of stars which were not used to
generate the polynomial.  The re-mapping was performed using linear
interpolation and a triangle method to distribute the original pixel
values onto the new map.  The method preserved surface brightness and
has been tested to ensure that in the case of a fractional pixel shift
with no change in the shape or size of the pixel that the second
moments of the objects in the image are not shifted in a systematic
manner (although there is some noise added).

The re-mapped images then had their sky subtracted by fitting a low
order polynomial to the minima in the image.  A final image was
produced by averaging the sky-subtracted images with a sigma-clipping
algorithm to remove cosmic rays.  Photometry was determined by
comparing the magnitudes of the bright but unsaturated stars with
those in images of the field from the UH88$^{\prime \prime}$ telescope
and Isaac Newton telescope which were calibrated from Landolt
standards (Landolt \cite{Landolt}).
We assume a Galactic extinction 
$A(B) = 0.81\pm 0.12$ (Schlegel et al.~\cite{Schlegel}) and
derive $A(V) \approx 0.61$, $A(R) \approx 0.46$, $A(I) \approx 0.29$, 
and $A(K) \approx 0.07$, 
using the color conversions of Cardelli et al.~(\cite{Cardelli}).  
The 5$\sigma $ limiting magnitudes within a square 1\farcs 0 aperture are
$I$=25.1, $R$=27.1, and $V$=27.1.

\begin{figure}
\resizebox{\hsize}{!}{\rotatebox{-90}{\includegraphics{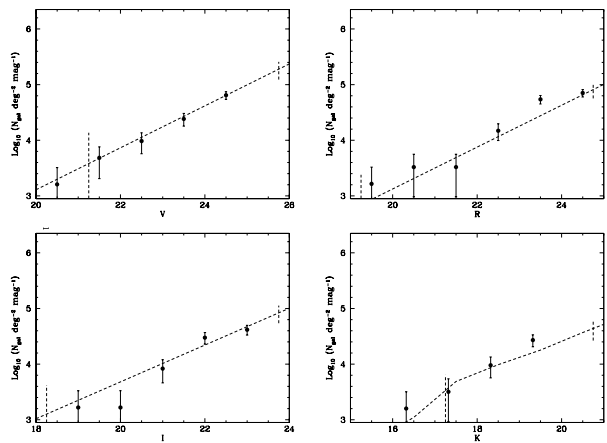}}}
\caption{Galaxy counts as a function of magnitude in the four
observed passbands.  All magnitudes 
are aperture magnitudes measured in an aperture of diameter 3.0 arcseconds,
which are close to total magnitudes except for very bright galaxies, where 
we apply a small aperture correction to the measured magnitudes so that they
approximate total magnitudes (which only happened for one foreground
galaxy).  The filled circles are the counts within $r=225$ kpc radius
from the QSO lens. The dashed lines represent typical mean background
counts (Mobasher \&\ Trentham \cite{Mobasher}, Wilson et
al.~\cite{Wilson}).  Approximate 
uncertainties in these lines are shown by the vertical dashed lines at two
representative magnitudes (Mobasher \&\ Trentham \cite{Mobasher}, Trentham
\cite{Trentham97}).}
\label{fig1}
\end{figure}

\begin{figure*}
\resizebox{\hsize}{!}{\rotatebox{90}{\includegraphics{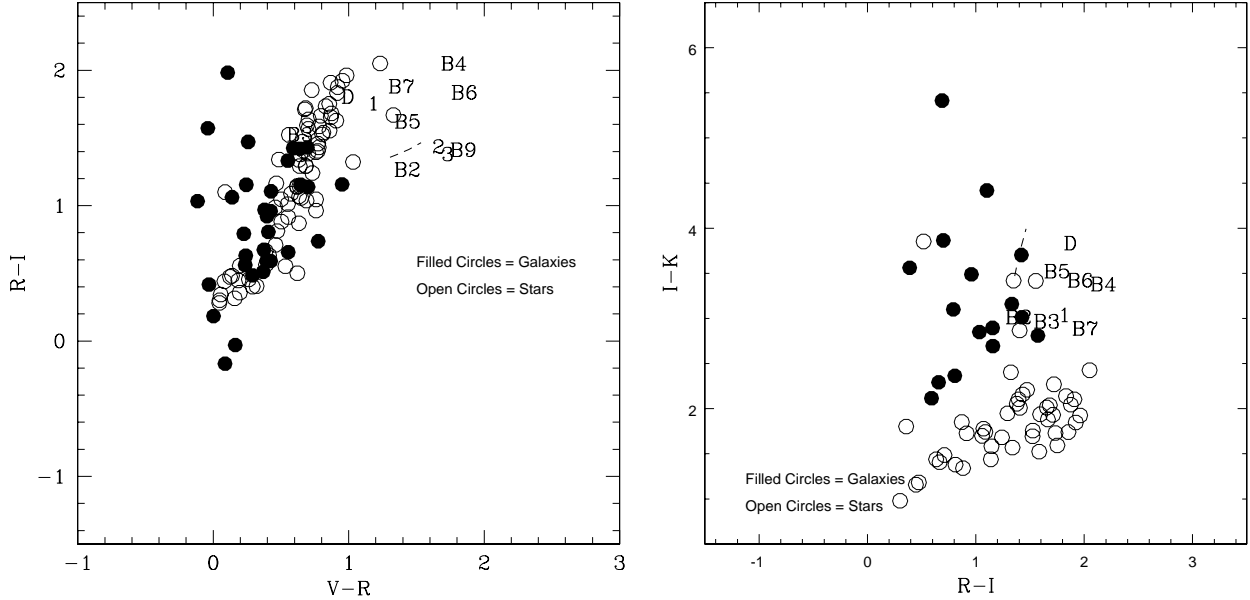}}}
\caption{Color-color diagrams for all objects with $K<20$ for the
$RIK$ plot and $R<24$ for the $VRI$ plot
which are less than 225 kpc from the QSO lens.  Symbol D
represents the positions for the four components of the lens galaxy. 
The symbols B\# represent the positions of the red galaxies in Table 1
of B99 where \# is the line number of the object in
the table.  The first object in their list is the same as object D.
The eighth and ninth objects were not detected in our $K$-band image
and the eighth object is believed to be a star from inspection of the
optical bands.  The symbols 1 through 3 represent new
candidate red galaxies (2 and 3 were not detected in our
$K$-band image either).  The dashed line is where one
expects passively evolving elliptical galaxies at $z=1$ to reside,
given the stellar population models of Kodama et al.~(\cite{Kodama}).}
\label{fig2}
\end{figure*}

\begin{figure}
\resizebox{\hsize}{!}{\includegraphics{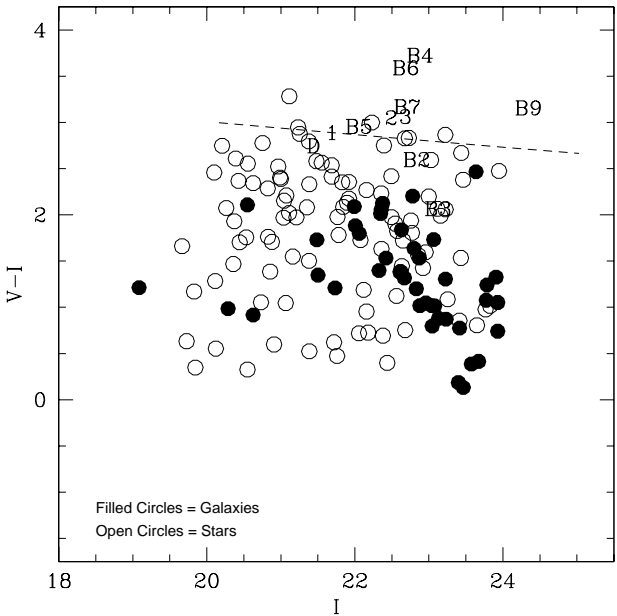}}
\caption{The $V-I$ vs.~$I$ color-magnitude diagram of all objects
less than 225 kpc from the QSO lens.  
The symbols have the same meaning as in Fig.~\ref{fig2}.
The dashed line is where one would
expect passively evolving elliptical galaxies at $z=1$ to reside, given the
stellar population models of Kodama et al.~(\cite{Kodama}).}
\label{fig3}
\end{figure}

We also obtained deep $K$-band images using IRCAM3 at UKIRT on 1998 July
19-20.  
Each image is 1\farcm 5 on a side with 0\farcs 286 pixels.
A 2$\times$2 mosaic of pointings resulted in a
3\arcmin$\times$3\arcmin \ field with 9720s exposure time in each
quadrant and an additional 3240s exposure time in the central 1\farcm
5$\times$1\farcm 5.
Individual exposures were 120s long and dithered by 8\farcs 0.  
The median seeing was about 0\farcs 8. 
Flat-field images were constructed using median-filtered dithered
object frames from the whole night and sky images were constructed using
sets of twenty-six 120s exposures.  
The individual frames were then
flat-fielded, sky-subtracted and combined using a clipping algorithm that
rejected bad pixels.     
Instrumental magnitudes were computed from observations of UKIRT faint
standards (Casali \&\ Hawarden \cite{Casali}), giving a photometric
accuracy in the zero-point of better than 2\%.
The 5$\sigma$ limiting magnitude within a 1\farcs 0 square aperture 
in the reduced image is $K$=21.3.

\section{Galaxy Counts and Color-Selection}

The IMCAT ({\it http://www.ifa.hawaii.edu/$\sim$kaiser/imcat})
{\it hfindpeaks} detection algorithm
was used to search for objects in the above $VRIK$ images. 
The objects were each inspected by eye in all images and spurious detections
(due to stellar diffraction peaks, for example) were eliminated from the
catalog of objects.  Objects were identified as stars or galaxies on the
basis of this inspection as well as using the measured half-light
radii and central surface brightnesses compared to the total
(large-aperture) magnitudes of the objects.  Those objects which were
classified as galaxies had their magnitudes remeasured
using the IRAF {\it apphot} package, following a local sky subtraction.
Uncertainties in these numbers are substantial at the faint end and
come from uncertainties in the local sky
due to scattered starlight, faint galaxy clustering, and Poisson noise.
For example, objects at $R>24$ typically have uncertainties of
at least
0.5 magnitudes in each color, measured by using apertures of various
sizes and radii around the objects to determine the local sky value.  

Fig.~\ref{fig1} shows the galaxy number counts as a function of
magnitude for this field in each filter.  
There is no obvious excess of galaxies above the background, as was seen
in the rich cluster MS1054$-$03 at $z=0.83$ (Trentham \&\ Mobasher
\cite{Trentham98}), but the background uncertainty
on these small scales is large.  Thus, the background
counts might be anomalously low, and a cluster at $z=1$ may still exist
despite the lack of an excess.  The most straightforward way to 
look for
such a cluster is to (i) look for dense groups of galaxies very close
to the lensing galaxy D, and (ii)
inspect color-color and color-magnitude diagrams.
These are presented in Fig.~\ref{fig2} and Fig.~\ref{fig3}.   

We identified one group of seven faint objects all between 19 and
27 kpc from galaxy D in the $R$-band image.  The brightest of these is
$R\approx 24$ and the rest around $R\approx 26$ and
all, except the brightest, were too faint to be detected in
the other filters.  The brightest has $R-I$ and $I-K$ colors of
1.57 and 2.81, which are somewhat bluer than those
of galaxy D (1.81 and 3.83).  From the colors it is unlikely that these
would be the brightest galaxies in a higher redshift cluster.  From
the observed number density of $24<R<26$ galaxies in this field, the
chances of projecting seven unrelated galaxies in such a small
area is only 0.14\%.  However, with the exception of the brightest
object, which is spatially extended and therefore, probably, a galaxy, 
we cannot determine if the objects are stars or galaxies.
Two other objects also around $R\approx 26$ are located
on the eastern side of galaxy D at roughly the same distance.  
Even assuming that these objects are galaxies and associated with
galaxy D, the proposed 8 keV cluster with galaxy D as the brightest 
cluster galaxy would still have a low surface density of cluster
galaxies near the BCG as compared to high-redshift clusters of similar
mass (both MS1137$+$66 and RXJ1716$+$67, $\sim$6 keV
clusters at $z\sim 0.8$, have roughly twice as many galaxies within 30
kpc of the BCG).  We note, however, that as MG2016$+$112 was
originally discovered as a medium-separation QSO lens, that this
deficiency of bright cluster galaxies near the BCG may be a selection
effect as a large number of massive galaxies would perturb the lensing
caustics sufficiently to preclude a double image of the QSO from
lensing by the BCG.

From Fig.~\ref{fig2} and Fig.~\ref{fig3}, one can find a number of red galaxies
having approximately the correct $I-K$ to be old stellar
populations at $z=1$ (Kodama et al.~\cite{Kodama}) within 225$h^{-1}$ kpc of
galaxy D.  Of these, eight are the red objects described
by B99, including galaxy D, and three are not found
in the B99 sample.  One of the B99 objects, the eighth in
their Table 1, we identified as a star.  Our revised count of the
total number of candidate members from color selection is thus eleven.
There are only fifteen galaxies in the image outside the selected
region with colors and magnitudes similar to the selected galaxies.
This gives an overdensity around galaxy D of $1160^{+440}_{-260}$\%\
compared to the rest of the image, and the probability this
is due to Poissonian fluctuations in the density is 3$\times 10^{-12}$.
If cluster members, as argued by B99, these are
presumably responsible for generating the metals seen at X-ray
wavelengths by H97.  Spectroscopic redshifts
would be needed to confirm this interpretation.

We note, however, that these galaxies are up to a magnitude redder
in $R-I$ than would be predicted for passively-evolving ellipticals.
This is probably due in part to photometric uncertainties, but
could also be due to small amounts of internal extinction. 
Note that at $z=1$ the $R-I$ color of a pure old stellar population
is anomalously red since at this redshift the 4000{\AA} break
falls exactly between these two passbands, which in turn means that 
this color is particularly susceptible to being made significantly 
different by internal extinction.  
In addition, galaxies bluer than the ones discussed by B99
could in principle also be members that are undergoing
current star formation.   Again, spectroscopic information
will be useful to investigate these possibilities further.

\section{Weak Lensing Analysis}
\subsection{Techniques}

The first step in the weak lensing analysis of the field is to detect
and measure the shapes of the background galaxies.  This was done
using the $R$-band image, which is of similar depth and resolution to
other Keck fields which have successfully reconstructed the surface mass
density of other high-redshift clusters (Clowe et al.~\cite{Clowe98},
\cite{Clowe00}).
Most of this analysis was done using the IMCAT software package.

Prior to using a detection algorithm on the image, we first subtracted
small-scale background fluctuations in the image, caused primarily by
stellar halos and scattered star light.   These fluctuations were
determined by smoothing an image of the local minima of the sky
values with a 5\arcsec \ Gaussian, and dividing by an image smoothed
on the same scale which contained the mask of the pixels detected as
minima.  In addition to removing any sky variations,
this routine typically removed extended wings of bright stars
in the image as well as any large, low surface brightness objects which
might have been present.  Performing this step was necessary as the
detection algorithm can get confused by rapid changes in the
sky background.

\begin{figure*}
\resizebox{\hsize}{!}{\includegraphics{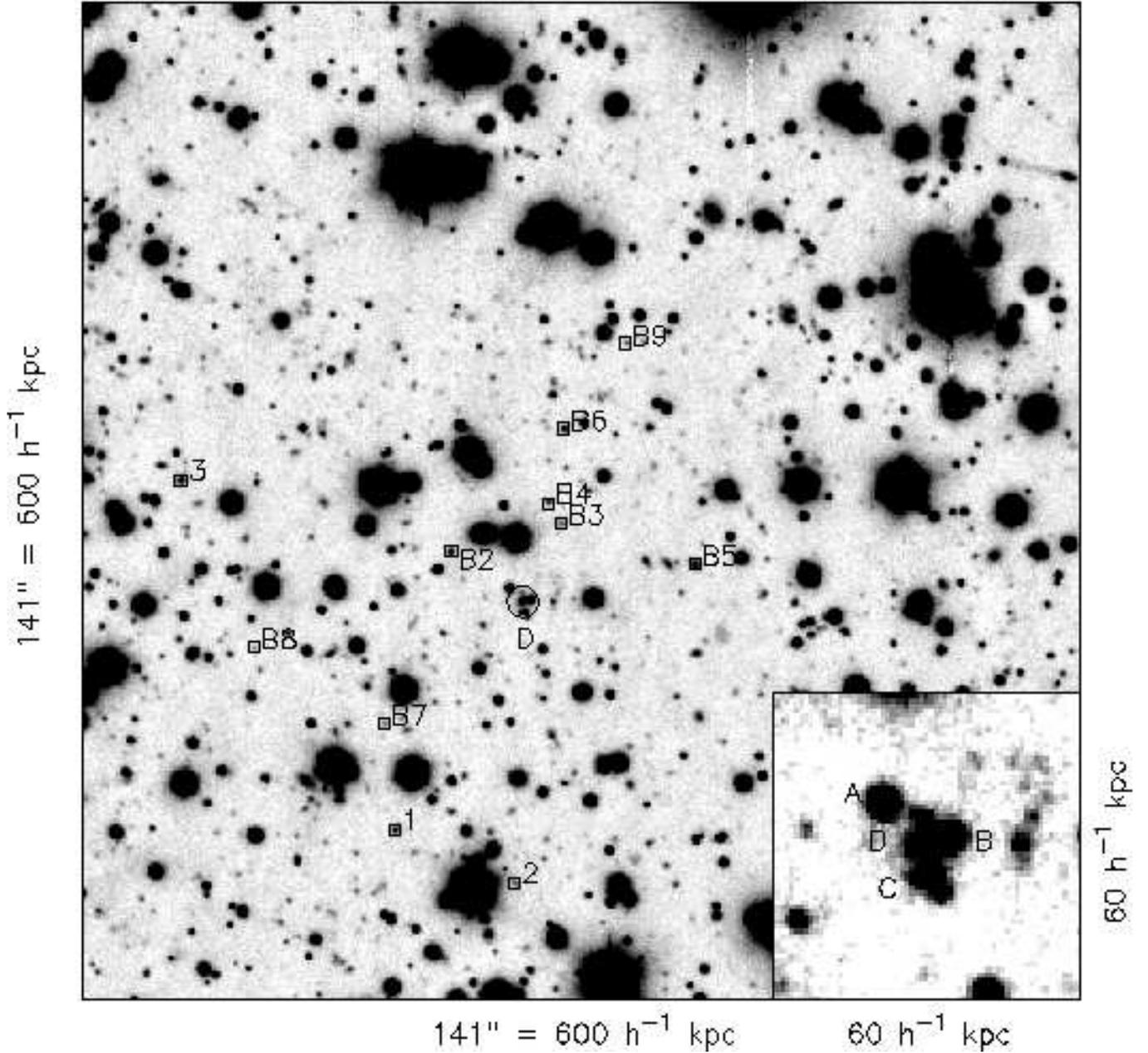}}
\caption{The central 600$h^{-1}$ kpc (141\arcsec ) of the $R$-band
image is shown above, 
plotted using a $\sqrt{\log}$ stretch, with north being up and east to
the left.  Objects identified as galaxies with colors consistent with
the QSO lensing galaxy have boxes drawn around them.  Inset in the
south-west corner is a close-up of the QSO lens, showing the cluster
of faint objects north-west of the lens system.}
\label{fig4}
\end{figure*}

To detect the faint galaxies we used a hierarchical peak-finding
algorithm which provided a position, luminosity, and size estimate for
each object (Kaiser et al.~\cite{Kaiser95}, hereafter KSB).  
Local sky levels
around each object were calculated after excluding all pixels within
three detection radii of an object in the catalog. The luminosity was
measured using a circular aperture equal to three times the smoothing
radius at which the object achieved maximum significance.  $I$ and $V$
magnitudes were measured using the same sky-subtraction technique and
the same aperture radius.  Noise peaks and merged objects were removed
from the catalog by requiring a minimum signal-to-noise at maximum detection
significance of 10, rejecting abnormally large and small objects,
rejecting objects with extremely high ellipticities, and by visual
inspection.  
The ellipticities of each object were measured using optical
polarizations $e_{\alpha} = \{I_{11}-I_{22},2I_{12}\}/(I_{11}+I_{22})$
formed from the quadrupole moments $I_{ij} = \int d^2\Theta 
W(\Theta)\Theta_i\Theta_jf(\Theta)$ where $f$ is the flux density and 
$W(\Theta)$ is a Gaussian weighting function of a scale equal to the 
smoothing radius at which the object was detected with maximum
significance (KSB).

\begin{figure}
\resizebox{\hsize}{!}{\includegraphics{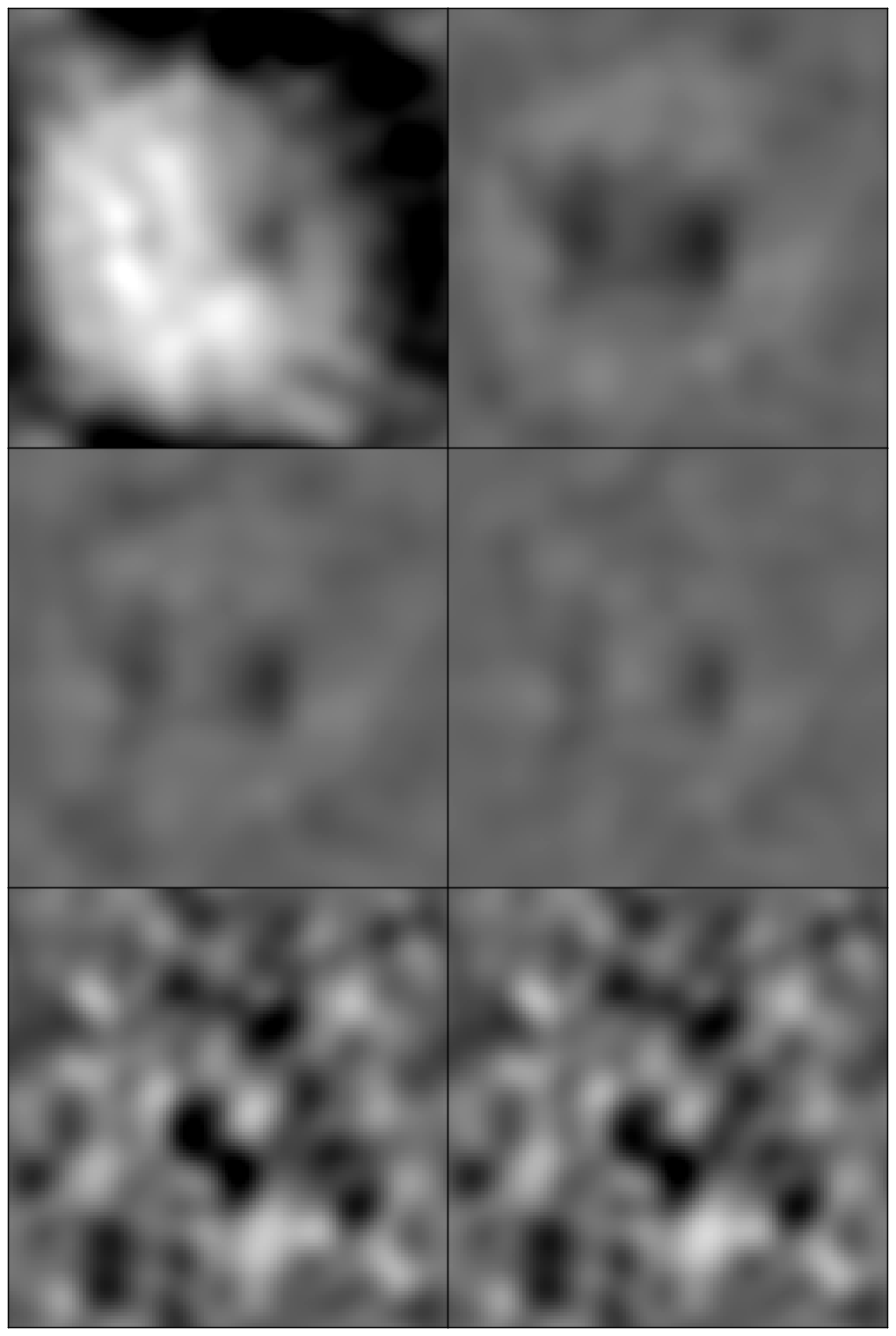}}
\caption{$\kappa$ measured from stars in the field.  The top-left
panel is the signal from the uncorrected ellipticities.  In the
top-right panel the ellipticities have had a bi-cubic fit subtracted.
The middle panels have the ellipticities corrected with a bi-pentic
and bi-septic fit for the left and right panels respectively.  The
bottom panels are mass reconstructions using the background galaxies
from the mass reconstruction in Fig.~\ref{fig7}, but applying a random
orientation to each galaxy while preserving the modulus of the
ellipticity, and thus is a good indication of the noise level seen in
the reconstructions from the intrinsic ellipticity distributions of
the background galaxies.  The left-hand image is the original
reconstruction and the right-hand image has the reconstruction for the
seventh-order corrected stellar ellipticities added to it.  The
greyscale is the same for all the panels, with black meaning positive
(false) mass.}
\label{fig5}
\end{figure}

The next step in the lensing analysis was to remove the PSF
anisotropies and smearing.  Due to the large number of stars in the
field (the final sample was 675 bright, unsaturated stars with no
neighbors within 10\arcsec), 
we were able to fit the stellar ellipticities across
the field with an arbitrarily high-order two-dimensional polynomial.  We
used third, fifth, and seventh order polynomials for the fits, and
subtracted the fit values from all detected objects via the method in
KSB.  We then applied the boost factor discussed below to the stellar
ellipticities, and used the boosted ellipticities to measure
any false signal which will be induced in the data by the residual PSF
anisotropies.  As can be seen in Fig.~\ref{fig5}, there is a small positive
signal in the center of the fields; however the magnitude of this
signal  is small compared to the noise caused by the intrinsic
ellipticity distribution of the background galaxies, and thus should 
not significantly affect any observed lensing signal.  The amplitude
of the false residual signal decreases somewhat with the increasing order of
the polynomial fit to the ellipticities, as can be seen in Fig.~\ref{fig5},
and thus we used the bi-septic polynomial fit to correct the
ellipticities of the galaxies.  This correction was done by
subtracting the value of the fit at the position of each galaxy from
the measured ellipticity, after scaling the fit value by
$P_{sm}/P^\star_{sm}$.  As defined in KSB (corrections in Hoekstra
et al.~\cite{Hoekstra}), $P_{sm}$ is a quasi-tensor which describes
how the ellipticity 
of an object changes under an applied anisotropy and the asterisk
denotes the typical value for a star in the image.

To correct for the reduction of the ellipticities from circular smearing by the
PSF, we calculated a boost factor for each galaxy 
$P_\gamma = P_{sh} - P_{sm}(P^\star_{sh}/P^\star_{sm})$ (Luppino \&
Kaiser \cite{Luppino}), where $P_{sh}$ is a quasi-tensor describing how the
ellipticity of an object changes with an applied shear (KSB), and
$P_{sm}$ and the asterisk are defined as above.  
Due to the large anisotropy in the
stars at the edges of the Keck field, we computed $P^\star_{sh}$ and
$P^\star_{sm}$ only in the central region with a diameter of one
thousand pixels, in which there
was no significant anisotropy.  As the individual values of $P_\gamma$
are quite noisy for faint galaxies, we fit $P_\gamma$ with a third
order polynomial as a function of
size and ellipticity.  Because of noise and a tendency of the
sky-subtraction algorithm to measure slightly too large a sky value
(owing to the presence of many faint, unresolved objects), very faint
galaxies tend to have artificially low, sometimes even negative,
$P_\gamma$ values.  We corrected for this by applying a cut to the
fitted $P_\gamma$ values such that any object with a detected size
smaller than that of the stars in the field was assigned a $P_\gamma$
equal to that of galaxies with detected sizes similar to that of
stars, 0.2 in this case.  We choose to apply this correction instead
of removing the objects from the catalog as the faintest galaxies
should be at the highest mean redshift, and therefore show the
greatest amount of gravitational shearing.  This artificial cutoff in
$P_\gamma$ is equivalent to applying a weight function in all subsequent
operations such that the galaxies with measured sizes smaller than
stellar are given a lower weight.

We then created an estimate of the shear for each galaxy 
$\hat{\gamma}_\alpha = e_\alpha/P_\gamma$.  Both the shear $\gamma$ 
and convergence $\kappa$ are second derivatives of the surface potential,
$\gamma _\alpha  = \hbox{$\,^1\!/_2$}\{\phi_{,11} - \phi_{,22}, 2\phi_{,12}\}$ and 
$\kappa = \Sigma/\Sigma_{crit} = \hbox{$\,^1\!/_2$}\nabla^2\phi$ where 
$\Sigma_{crit}^{-1} = 4\pi Gc^{-2}D_lD_{ls}D_s^{-1}$.  Thus, one can
create a two dimensional map of the convergence by Fourier transforming
shear estimates, multiplying by the appropriate conversion factors,
and retransforming (Kaiser \&\ Squires \cite{Kaiser93}).   The surface
density can then be
extracted from the convergence, providing one knows both the redshift
of the lens and the redshift distribution of the background
galaxies.  The intrinsic ellipticities of the galaxies cannot be
removed from the shear estimate, and are the dominant source of noise
in the analysis.

\begin{figure}
\resizebox{\hsize}{!}{\includegraphics{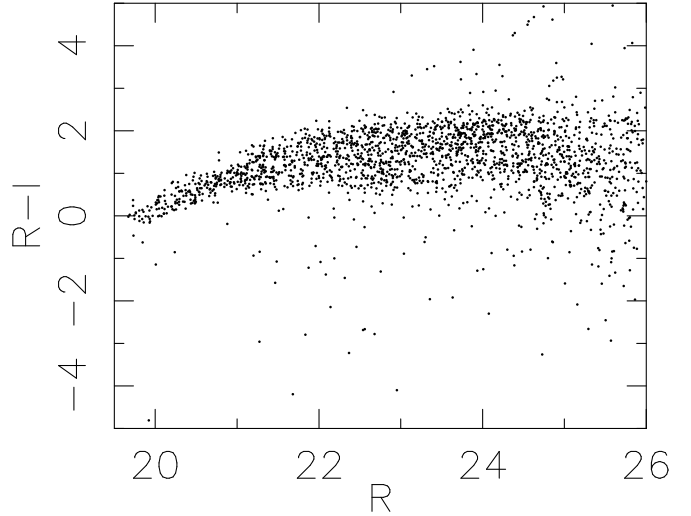}}
\caption{$R-I$ vs.~$R$ color-magnitude diagram of all objects
detected with a half-light radius the same as the bright
stars.  For $R<24$, this will be almost entirely stars while for
$R>24$ small galaxies will be included in the objects.}
\label{fig6}
\end{figure}

\subsection{Results}
Previous work on lensing by $z\sim 0.8$ clusters has shown that to
maximize the signal-to-noise of the lensing one must use faint blue
galaxies (Luppino \&\ Kaiser \cite{Luppino}; Clowe et
al.~\cite{Clowe98}, \cite{Clowe00}) as these are 
presumably at higher redshifts than the brighter and redder galaxies,
and thus have been subjected to a stronger shear.  Also, because of
the large number of stars in this field, one must also apply a
selection in color to exclude the majority of the stellar population,
which still has a significant contribution at the faintest observable
magnitudes (as seen in Fig.~\ref{fig6}).
As a result we selected as our background sample all galaxies with
$24.20<R<26.70$ and $R-I<0.35$ ($23.64<R<26.14$ and $R-I<0.55$ after
correcting for Galactic extinction).  This resulted in a catalog of 910
galaxies ($\sim 17$ per sq. arcminute), which is less than half the
counts expected from the color and magnitude cuts 
because of incompleteness at the faint end and loss of
area due to extended stellar halos.  The rms ellipticity of this
galaxy sample, after the corrections for psf effects described above,
is $\sim 0.4$, which combined with the number counts gives an rms
shear in a square arcminute area of $\sim 0.1$.

\begin{figure}
\resizebox{\hsize}{!}{\includegraphics{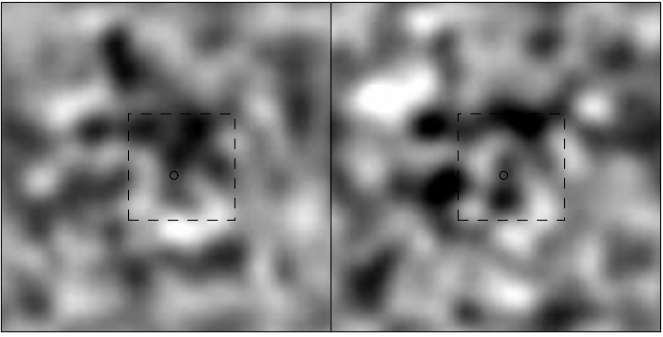}}
\caption{The weak lensing $\kappa$ reconstructions, using the KS93
algorithm, is shown above.  Positive $\kappa$ peaks are black, and the
dashed square indicates the area of the reconstruction corresponding
to the image shown in Fig.~\ref{fig4}.  The circle indicates the position of
the QSO lens.  The left-hand map uses objects detected in the $R$-band
image, and the right-hand map uses objects detected in the combined
$R+V$ image.}
\label{fig7}
\end{figure}

The weak lensing mass reconstruction from the background galaxy
catalog is shown in Fig.~\ref{fig7}.  This was created using the KS93
algorithm (Kaiser \&\ Squires \cite{Kaiser93}), which uses the Fourier transform
method described earlier to convert shear to convergence.
This algorithm can only determine $\kappa $ to an
unknown additive constant and suffers from biasing at the edges of the
frame, but the intrinsic ellipticity distribution of the galaxies is
translated into white noise across the field.  

As can be seen in Fig.~\ref{fig7}, we do detect an overdensity of mass in the
field, although its center is 60\arcsec north and 24\arcsec east of
the position of the QSO lens.  The rest of the peaks seen in the
reconstruction are at the level of the expected noise due to the low number
density of background galaxies.  One way to check that
the peak is real and not caused by a small number of high ellipticity
galaxies is to randomly divide the background galaxy catalog in two
and see if the peak is present in both of the sub-catalogs.  We did
this ten times, and in all twenty sub-catalogs we detect the mass peak
albeit with changes in the shape, position, and amplitude as expected
by the increased noise in the reconstruction due to the lower
background galaxy number density.  The other peaks in the field tended
to vary in amplitude and often disappear in the sub-catalog
reconstructions, which suggests that they are caused primarily by a
handful of galaxies and not a real lensing signal.  

We also performed
a lensing analysis using an image made from both the $R$ and $V$
images added together.  This allowed us to detect $\sim 20\%$ more
galaxies using the same signal-to-noise, $R$ magnitude, and $R-I$
color cuts as given above.  The lensing reconstruction for this is
also shown in Fig.~\ref{fig7}.  The extra galaxies and change in the
noise properties of the original galaxies results in only a small
change in the amplitude of the main peak
but a much larger change in the smaller, presumably noise, peaks.
We also attempted the same using $R$ and $I$, and $V$, $R$, and $I$
images added together, but the higher sky noise in the $I$ frames
resulted in a severe decrease in the number of faint galaxies
detected.  For the rest of the results we will quote the numbers taken
from just the $R$ image, but all of the results from the $R+V$ image
agree within errors to those of the $R$ image.

\begin{figure}
\resizebox{\hsize}{!}{\includegraphics{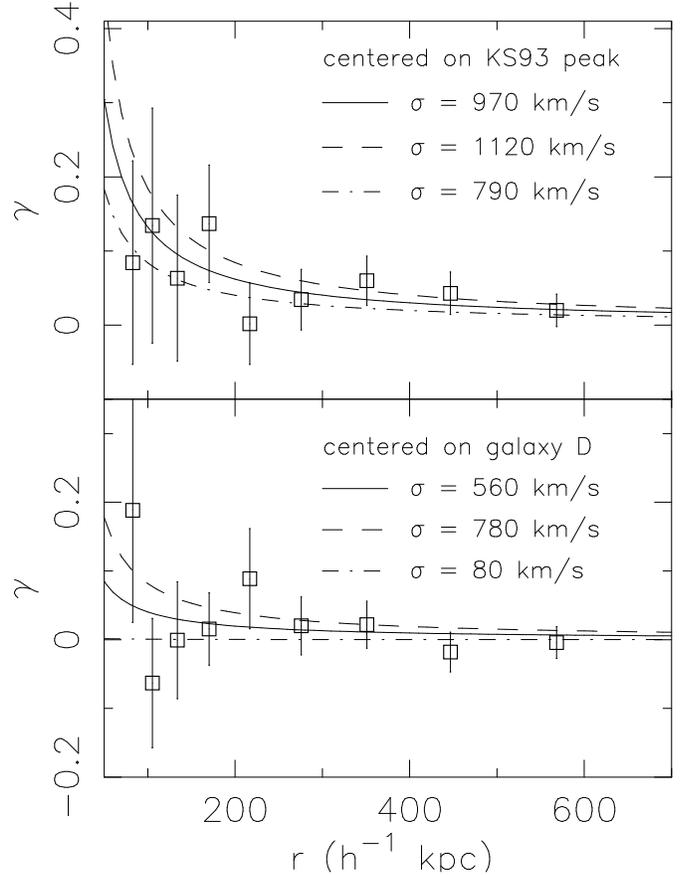}}
\caption{The radial shear profiles centered on the peak of the KS93
$\kappa$ reconstruction (top) and the QSO lens (bottom).  
The solid lines show the best fit singular isothermal sphere
profile assuming $z_{bg} = 1.5$ and $z_{lens} = 1$ and the dashed
lines are the 1$\sigma$ limits determined by $\delta \chi^2$ of the
fits.}
\label{fig8}
\end{figure}

To better quantify the strength and significance of the observed
overdensity, as well as to detect any low-significance signal at the
position of the QSO lens, we computed the azimuthally-averaged
tangential shear of the background galaxies as a function of
radius from a given position.  We can then fit these radial shear
profiles with various models and measure the $\delta \chi ^2$ between
the best fit model and a zero shear model to determine the
significance of the peaks.  The radial shear profiles centered on the
peak seen in the mass reconstruction and on the QSO lens are given in
Fig.~\ref{fig8}, as well as the best fit isothermal sphere models.  For
these models we assumed a lens redshift of 1 and background galaxy
redshift of 1.5.  The $z_{bg}$ = 1.5 assumption was derived by
calculating 
\begin{equation}
\bar{\Sigma }_{crit}(z_bg) = {\int n(z) dz\over \int 
{n(z)\over \Sigma _{crit}(z)} dz}
\end{equation}
where $n(z)$ is from the HDF south photometric redshift survey of
Fontana et al.~(\cite{Fontana}) using the same magnitude and color cuts as used
in the data, after correction for Galactic extinction.  
Altering these assumptions will change the velocity
dispersion, and thus the mass, of the lens model, but does not
significantly change the $\chi ^2$ or the significance of the peak.

\begin{figure}
\resizebox{\hsize}{!}{\includegraphics{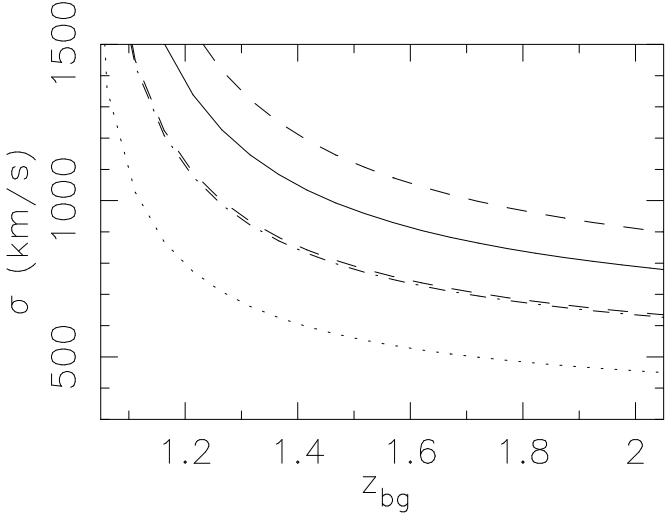}}
\caption{The best fit singular isothermal sphere
velocity dispersion to the observed shear when centered on the KS93
mass peak (solid line) and QSO lens (dotted line) as a function of
mean background galaxy redshift, assuming the lens is at $z=1.01$.
The dashed and dash-dotted lines
give the 1$\sigma$ limits for the two centers, respectively.}
\label{fig9}
\end{figure}

The best fit isothermal sphere model for the observed peak, centered
on the peak in the KS93 reconstruction, has $v
=$ 970 km/s, $\chi ^2/(n-1) =$ 1.02, and $\delta \chi ^2$(0 km/s $-$ 970
km/s) = 8.95, and thus is at nearly 3$\sigma $ significance.  The
highest significance in the field using this statistic occurs 5\farcs
6 east and 14\farcs 2 north of the KS93 peak (18\farcs 3 west and
71\farcs 6 north of the QSO lens), with $\sigma =$ 1040 km/s, $\chi
^2/(n-1) =$ 0.62, and $\delta \chi ^2$(0 km/s - 1040 km/s) = 9.23.
The difference between the maxima in the two techniques is a result of
the KS93 reconstruction using all of the galaxies in the frame while
the tangential shear uses only those between 17\farcs 2 and 150\farcs
5 from the central position, and that the contributions from the
galaxies are weighted in different manners.  The best
fit isothermal sphere model when the shear profile is centered on the
QSO lens has $\sigma = $ 560 km/s, $\chi ^2/(n-1) =$ 1.13, and $\delta
\chi ^2$(0 km/s $-$ 560 km/s) = 1.05, and therefore has only
$1\sigma$ significance.  The best fit velocity dispersion, as well as
the one-$\sigma$ deviations, as a function of background
galaxy redshift is given in Fig.~\ref{fig9}.

To test the significances of the peaks given by the $\delta \chi ^2$
of the isothermal sphere fits we performed Monte Carlo simulations in
which we kept the positions and moduli of the ellipticities of the background
galaxies fixed, but applied a random spin to their orientations.  We
then measured the best isothermal sphere fits to the tangential shear 
profiles using the central position from the KS93 reconstruction.
The resulting best fit isothermal sphere models were a Gaussian
distribution around $v^2 =$ 0 km$^2$/s$^2$.  Only 128 of the 100,000
realizations resulted in a best fit model with $\sigma \geq $ 970
km/s, which corresponds to a significance level of 3$\sigma $, in
agreement with the $\delta \chi ^2$ significance level.

To determine the significance of not detecting a peak at the position
of the QSO lens we performed a similar process to the above.  We again
kept the positions and moduli of the ellipticities of the background galaxies
fixed and randomized their orientations, but then we also applied a
shear to the background galaxies.  The applied shear was calculated
for each galaxy's position by using a 1000 km/s isothermal sphere
centered on the QSO lens with
$z_{lens} = 1$ and $z_{bg} = 1.5$.  Only 5031 of the 100,000
realizations resulted in a best fit model with $\sigma \leq$ 560 km/s,
which corresponds to $2.0\sigma $ significance.

\section{Conclusions}

We have detected in $VRIK$ images ten galaxies with colors consistent
with the lensing galaxy of the MG2016$+$112 QSO lens within
$225h^{-1}$ kpc of the lensing galaxy.  This represents an increase in
number density of $1160^{+440}_{-260}$\%\ (1$\sigma$ error on the
background density) compared to galaxies of
similar colors and magnitudes in the rest of the image.  We also find
nine faint objects, seven of them grouped in a small area, within
$30h^{-1}$ kpc of the lens galaxy D.  Due to their faintness, we are
unable to measure colors or determine if they are galaxies or stars
except for the brightest of the nine which is a galaxy and somewhat
bluer than galaxy D.  Even assuming all of these galaxies are
associated with the lens galaxy D, they would total a small fraction of the
galaxy counts associated with the clusters MS1137$+$66 and
RXJ1716$+$67, both at $z\sim 0.8$ and of similar X-ray properties as
those measured for AXJ2019$+$112 (Donahue et al.~\cite{Donahue}; Henry et al.~
\cite{Henry}; H97), using the same radial, magnitude, and color
cuts adjusted to the cluster BCGs and redshifts.

\begin{figure}
\resizebox{\hsize}{!}{\includegraphics{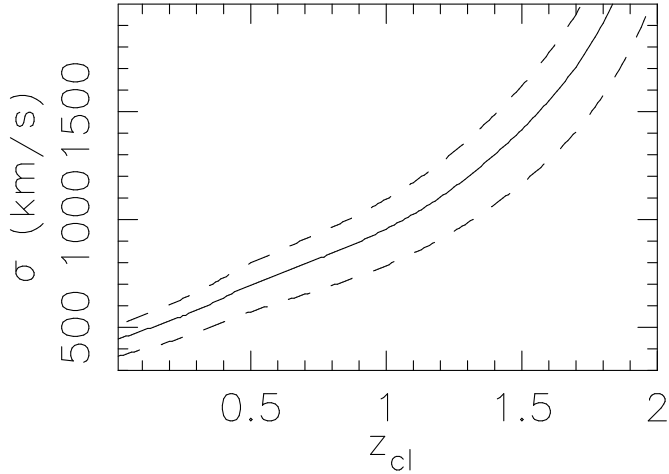}}
\caption{The best fit singular isothermal sphere
velocity dispersion to the observed shear centered on the KS93 mass
reconstruction peak as a function of lens redshift.  The background
galaxies are assumed to have the same redshift distribution as those
of similar magnitude and redshift in the Fontana et al.~(\cite{Fontana}) HDF-S
photometric redshift catalog.  The dashed lines give the $1\sigma$
limits to the fits.}
\label{fig10}
\end{figure}

We have also made a mass reconstruction of the field from faint blue
galaxies.  We find a peak of comparable mass to that given by the
X-ray observations, but the center of the peak is over an arcminute
north-west of the lens galaxy D (272$h^{-1}$ kpc at $z=1$).  This peak
has a best fit singular isothermal sphere model of $\sigma =$ 970
km/s and is significant at $3\sigma$.  We can rule out a 1000 km/s
singular isothermal sphere, the mass expected from the X-ray
observations, centered on the lens galaxy D at 2.0$\sigma$
significance.  We do not see any evidence for a group of similarly
colored galaxies in the vicinity of the observed mass peak, but this
region is highly obscured by bright stars.  
We plot in Fig.~\ref{fig10} the velocity dispersion of a singular isothermal
sphere needed to provide the observed lensing signal as a function of
redshift of the lens.  

Finally, we note that the 1$\sigma$ upper mass limit centered on
galaxy D from weak lensing is near the 1$\sigma$
lower mass limit from the X-ray data.  Thus the best measure of the
mass of the cluster from the two data sets would give a velocity
dispersion of $\sigma \sim $800 km/s.

\begin{acknowledgements}

We wish to thank Genevieve Soucail, Jean-Paul Kneib, Narciso Benitez,
Tom Broadhurst, and Piero Rosati for providing information and data on this
field prior to publication.  We also wish to thank Tadayuki Kodama and
Peter Schneider for their advice and help.  Some of the data presented
herein were obtained at the W.M. Keck Observatory, which is operated
as a scientific partnership among the California Institute of 
Technology, the University of California and the National Aeronautics
and Space Administration.  The Observatory was made 
possible by the generous financial support of the W.M. Keck Foundation.
The United Kingdom Infrared Telescope is operated by the Joint
Astronomy Centre on behalf of the U.K. Partical Physics and Astronomy
Research Council.
We acknowledge assistance from the Isaac Newton Group Service Programme
in providing photometric zero points for the optical data, from the Isaac
Newton Telescope on La Palma.
DC acknowledges the ``Sonderforschungsbereich 375-95 f\"ur
Astro--Teil\-chen\-phy\-sik" der Deutschen
For\-schungs\-ge\-mein\-schaft for finacial support.  NT acknowldedges
the PPARC for financial support.
This work was supported by the TMR Network ``Gravitational Lensing:
New Constraints on Cosmology and the Distribution of Dark Matter'' of
the EC under contract No. ERBFMRX-CT97-0172.

\end{acknowledgements}

\end{document}